\documentclass[sigconf]{acmart}

\copyrightyear{2026}
\acmYear{2026}
\setcopyright{cc}
\setcctype{by}
\acmConference[DAC '26]{63rd ACM/IEEE Design Automation Conference}{July 26--29, 2026}{Long Beach, CA, USA}
\acmBooktitle{63rd ACM/IEEE Design Automation Conference (DAC '26), July 26--29, 2026, Long Beach, CA, USA}
\acmDOI{10.1145/3770743.3803985}
\acmISBN{979-8-4007-2254-7/2026/07}

\usepackage[dvipsnames]{xcolor}
\usepackage{algorithmic}
\usepackage{graphicx}
\usepackage{textcomp}
\usepackage{xcolor}
\usepackage{multirow}
\def\BibTeX{{\rm B\kern-.05em{\sc i\kern-.025em b}\kern-.08em
    T\kern-.1667em\lower.7ex\hbox{E}\kern-.125emX}}

\usepackage{subfiles}
\usepackage{booktabs}
\usepackage{subfig}
\usepackage{siunitx}
\usepackage{enumitem}

\usepackage[table]{xcolor} 
\usepackage{array}

\newcolumntype{R}{>{\columncolor{green!20}}r}

\begin{document}


\title[Duplication-Aware Retiming and Cell Interface Redesign for Superconductor Circuit Minimization]{Duplication-Aware Retiming and Cell Interface Redesign for Superconductor Circuit Minimization}

\author{Panagiotis Papanikolaou}
\email{ppapanikolao@wisc.edu}
\affiliation{%
  \institution{University of Wisconsin - Madison}
  \city{Madison}
  \state{Wisconsin}
  \country{USA}
}
\author{Alex \\ Vanasse}
\email{alvanasse@wisc.edu}
\affiliation{%
  \institution{University of Wisconsin - Madison}
  \city{Madison}
  \state{Wisconsin}
  \country{USA}
}

\author{Haoran \\ Jin}
\email{allenjin@umich.edu}
\affiliation{%
  \institution{University of Michigan}
  \city{Ann Arbor}
  \state{Michigan}
  \country{USA}
}

\author{Georgios \\ Tzimpragos}
\email{tzimpragos@wisc.edu}
\affiliation{%
  \institution{University of Wisconsin - Madison}
  \city{Madison}
  \state{Wisconsin}
  \country{USA}
}
\author{Jennifer \\ Volk}
\email{jennifer.volk@wisc.edu}
\affiliation{%
  \institution{University of Wisconsin - Madison}
  \city{Madison}
  \state{Wisconsin}
  \country{USA}
}

\settopmatter{authorsperrow=5}

\begin{abstract}
Superconductor electronics have increasingly shifted away from RSFQ and its variants toward logic families that eliminate explicit gate-level clocking. While this transition enables simpler circuits and more efficient architectures, it also introduces an implicit reliance on dual-rail codes, resulting in inherent gate duplication. This work presents a duplication-aware retiming methodology for Josephson junction (JJ) count minimization, co-optimizing register placement and polarity assignment. The approach applies beyond SFQ to any monotonic circuit. We further identify cell interfaces—specifically, interconnect drivers, receivers, and fanout (FO) elements—as dominant contributors to JJ count in each cell. A new amplifier design is introduced to reduce these costs, integrated within existing SFQ cells, experimentally verified, and characterized to form a new SFQ cell library. Our results demonstrate a 63-71\% JJ count reduction in single-cycle implementations and 41-66\% reduction in multi-cycle implementations compared to the prior state-of-the-art. The latter establishes a new Pareto frontier, achieving both shorter critical paths and lower JJ counts than the best-to-date single-cycle implementations.

\end{abstract}


\maketitle
\renewcommand{\shortauthors}{Panagiotis Papanikolaou, Alex Vanasse, Haoran Jin, Georgios Tzimpragos, and Jennifer Volk}


\section{Introduction} 
\label{sec:intro}

From advanced sensors~\cite{sensors} to high-performance and quantum computing systems~\cite{superconducting-qubits}, superconductor electronics (SCE) open pathways for next-generation technologies. The use of superconducting materials eliminates resistive wiring, a key drawback in CMOS designs, and the Josephson effect enables high-speed, energy-efficient switching via Josephson junctions (JJs) rather than transistors.

JJ-based circuits are far from new~\cite{squid}. However, the integration density of superconducting devices remains limited. Current SCE technologies can reach 1-10 million JJs per cm$^2$~\cite{tolpygo2022scalability,tolpygo2023progress}, whereas advanced CMOS processes integrate more than 10 billion transistors per cm$^2$. Although planned fabrication advances~\cite{tolpygo2024development} promise improvements, they demand significant time and capital investment and are unlikely to close the integration density gap with CMOS.

We propose a complementary path forward grounded in design automation and circuit advancements. Our objective is to maximize functionality per unit area---that is, to deliver more complex behavior using fewer JJs. We refer to this as functional density to distinguish it from the integration density discussed above.

Surveying the SCE design automation literature reveals substantial work on improving logic synthesis for rapid single-flux-quantum (RSFQ) circuits~\cite{coldflux,qpalace,pbmap}. RSFQ emerged in the mid-1980s~\cite{rsfq,oleg-rsfq} and, together with its variants (e.g., ERSFQ~\cite{ersfq}, eSFQ~\cite{esfq}), dominated the field for over three decades. In the past ten years, while RSFQ remains in use, the SCE community has increasingly shifted focus toward logic families that eliminate the need for explicit gate-level clocking. Prominent examples from both industry and academia include reciprocal quantum logic (RQL, 2011)~\cite{rql}, dynamic SFQ (DSFQ, 2019)~\cite{dsfq}, alternating SFQ (xSFQ, 2021)~\cite{xsfq}, and pulse-conserving logic (PCL, 2023)~\cite{pcl}.

The mechanisms by which these logic families circumvent explicit clocking differ. For instance, RQL and PCL rely on multi-phase AC biasing, which enables circuit operation and performs resetting as the phase evolves. DSFQ gates operate logically asynchronously and use resistors to dissipate stored magnetic fluxons in their SQUIDs for resetting. xSFQ also evaluates asynchronously, like DSFQ, but employs alternating codes for resetting, eliminating the need for resistors or multi-phase AC biasing. Despite these differences, all these approaches share two key features: (a) to avoid signal races and clocking at inverters, they are effectively dual-rail; and (b) like RSFQ and its variants, they generally assume the use of passive transmission lines (PTLs) for inter-gate interconnection~\cite{interconnect}.


Recent work on dual-rail SCE logic synthesis addressed point (a) by leveraging similarities to domino logic~\cite{xsfq-dac}. This method reduced inherent dual-rail duplication through a graph-coloring algorithm that assigns a single polarity at synchronous interfaces and propagates these choices backward to remove redundancies. The result: over 4$\times$ lower JJ count over the prior state of the art~\cite{pbmap,qseq} across combinational and sequential benchmarks.


In this work, we introduce new strategies to further reduce gate duplication while addressing point (b). Our key contributions are: \textbf{(1)} a duplication-aware retiming method that minimizes cell and JJ count by optimizing the tradeoff between register placement and de-duplication via an ILP; and \textbf{(2)} a new integrated xSFQ-amplifier cell library that reduces JJ count for fanout and PTL driving by 1.7-3.1$\times$ (FO2) and 2.8-6.0$\times$ (FO4). To our knowledge, (1) is the first retiming solution for monotonic circuits jointly optimizing sequential and combinational logic, and (2) yields the most compact SFQ cell library to date, accounting for all interfacing costs.

Evaluation highlights include:
\begin{itemize}[leftmargin=10pt]
    \item Geometric mean reduction of 7–15\% of JJs across pipelines and timing targets for the duplication-aware algorithm of point (1) alone, with up to 22\% improvement on individual benchmarks.
    \item JJ count reduction of 63–71\% (2.7–3.4$\times$) on single-cycle implementations for the redesigned xSFQ-amplifier cells of point (2) alone, compared to prior work~\cite{xsfq-dac}. A prototype cell was fabricated and experimentally verified.
    \item JJ count reduction of 41–66\% (1.7–2.9$\times$) on multi-cycle implementations when combining points (1) and (2), with the narrower range reflecting the tradeoff between throughput (pipeline depth) and JJ resources (e.g., registers). These designs establish a new Pareto frontier, achieving both shorter critical paths and lower JJ count than the best-to-date single-cycle implementations~\cite{xsfq-dac}. 
    
\end{itemize}

\section{Background} 
\label{sec:back}
\label{sec:2}
\subsection{SCE Digital Logic}
\label{sec:digital-logic}
In SCE circuits, information is typically encoded as picosecond-wide, millivolt-amplitude SFQ pulses that propagate ballistically between gates. Unlike CMOS, where logic levels can be directly aligned, synchronizing these ultrashort pulses for logic operations is practically impossible. As a result, SFQ gates must be capable of retaining pulses, effectively storing the state of their arrival or absence. In other words, each gate can be viewed as a small Mealy machine: the arrival of a pulse within a predefined interval represents a logic 1 and triggers a state transition, whereas its absence represents a logic 0 and leaves the gate state unchanged.

The very notion of an interval requires the presence of start and end events. In RSFQ and RSFQ-like implementations, this timing is enforced through an explicit SFQ clock signal distributed to every gate. More recent SFQ logic families eliminate gate-level clocks and support only monotonic functions, such as AND and OR. This is because pulse encoding permits only $0 \rightarrow 1$ transitions within a cycle. Without native inversion, these families rely on dual-rail encoding to implement arbitrary Boolean logic, making some level of gate duplication unavoidable.

\begin{figure} [b]
    \centering
    \includegraphics[width=0.85\linewidth]{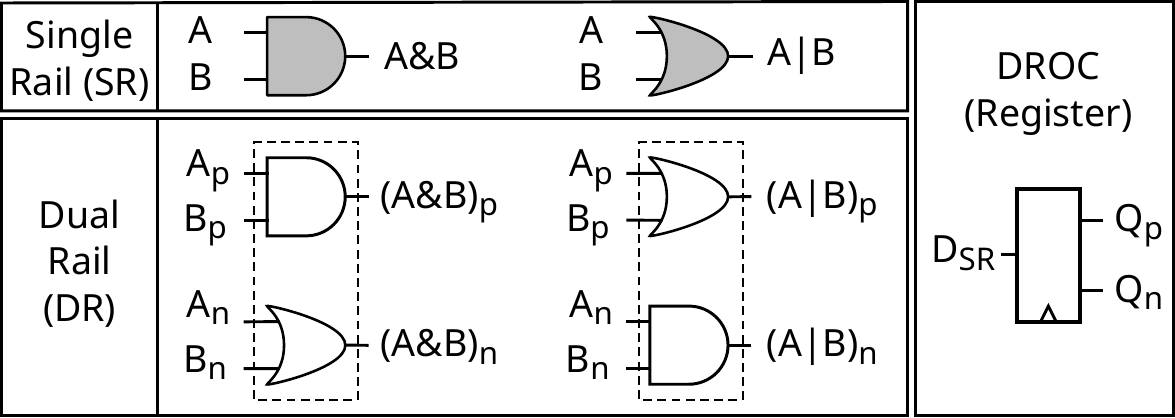}
    \caption{Single-rail gates (SR, shaded) that must be duplicated in dual rail are mapped to an FA–LA (OR–AND) pair. The DROC register accepts an SR input of either polarity (subscripts $p$ or $n$) and produces the corresponding DR outputs for the next stage.}
    \label{fig:duplication}
\end{figure}

\subsection{xSFQ Logic Synthesis}
\label{sec:synthesis}
We center our analysis on xSFQ circuits because of their simplicity, EDA-flow support~\cite{xsfq-dac}, architectural flexibility~\cite{xsfq}, and compatibility with low-energy biasing techniques~\cite{xesfq, lupo2025digital}.


\subsubsection{xSFQ basics}
xSFQ uses only two logic primitives and a storage element: First Arrival (FA), Last Arrival (LA), and a Destructive Read-Out cell with Complementary outputs (DROC)~\cite{xsfq-dac}. DROC is the only synchronous component and functions similarly to a flip-flop with complementary outputs. FA and LA can be interpreted as the single-rail equivalents of OR and AND gates, respectively.


To implement dual-rail OR and AND functions, each logic gate requires a pair of FA-LA cells. Without reducing gate duplication, the circuit ends up using twice as many FA and LA cells as the original number of single-rail gates. Figure~\ref{fig:duplication} provides an illustration.

\subsubsection{Polarity assignment}
Because DROC cells generate complementary outputs, duplication can be reduced by propagating only one polarity (positive or negative, denoted by subscripts $p$ or $n$) of each dual-rail signal to the DROC input (Figure~\ref{fig:duplication}). Each combinational block feeding a DROC cell therefore needs only a single output polarity, allowing the logic that produces the unused polarity to be removed. Removable gates are identified by traversing the netlist backward from the DROC cell, terminating either at the inputs to that stage or at nodes where both polarities are consumed. If a gate must be duplicated, its fanin gates must be, as well.

\begin{figure} [t]
    \centering
    \includegraphics[width=0.8\linewidth]{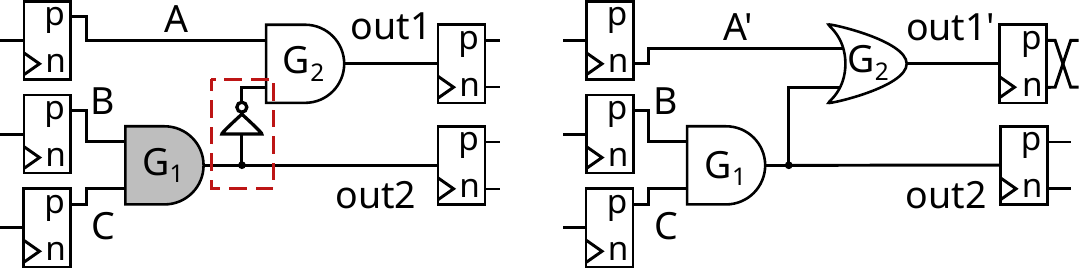}
    \caption{Left---$G_2$ needs no duplication because a single polarity ($p$) allows the DROC to generate both rails. $G_1$, however, must be duplicated because its two output paths have opposite polarities ($n$ above and $p$ below). Right---Applying De Morgan's rule at $G_2$ makes both paths positive, eliminating duplication in both $G_1$ and $G_2$. }
    \label{fig:demorgan}
\end{figure}

Figure~\ref{fig:demorgan} illustrates this process. When both outputs are positive, gate $G_2$ requires no duplication. In contrast, gate $G_1$ must be duplicated because it drives paths with conflicting polarity ($p$ vs. $n$). Applying De Morgan's rule at $G_2$ makes $G_1$'s output paths polarity-consistent ($p$), removing duplication from both $G_1$ and $G_2$.

A closer look at this process reveals two key factors for minimizing duplication: (a) optimal output polarity assignment and (b) strategic placement of DROC cells within combinational logic---e.g., via retiming---to absorb inversions. Prior xSFQ synthesis work~\cite{xsfq-dac} has addressed the first factor using techniques adapted from domino logic~\cite{domino-graph}. The second factor is the focus of this paper.

\section{Duplication-Aware Retiming} 
\label{sec:retiming}

\label{sec:3}
Traditional minimum-area retiming assumes register placement does not affect combinational logic and focuses only on minimizing register counts while meeting the critical-path delay~\cite{timing-book}. This assumption does not hold in dual-rail circuits, where the placement of registers (i.e., DROC cells) directly influences gate duplication.

\begin{figure} [t]
    \centering
    \includegraphics[width=0.85\linewidth]{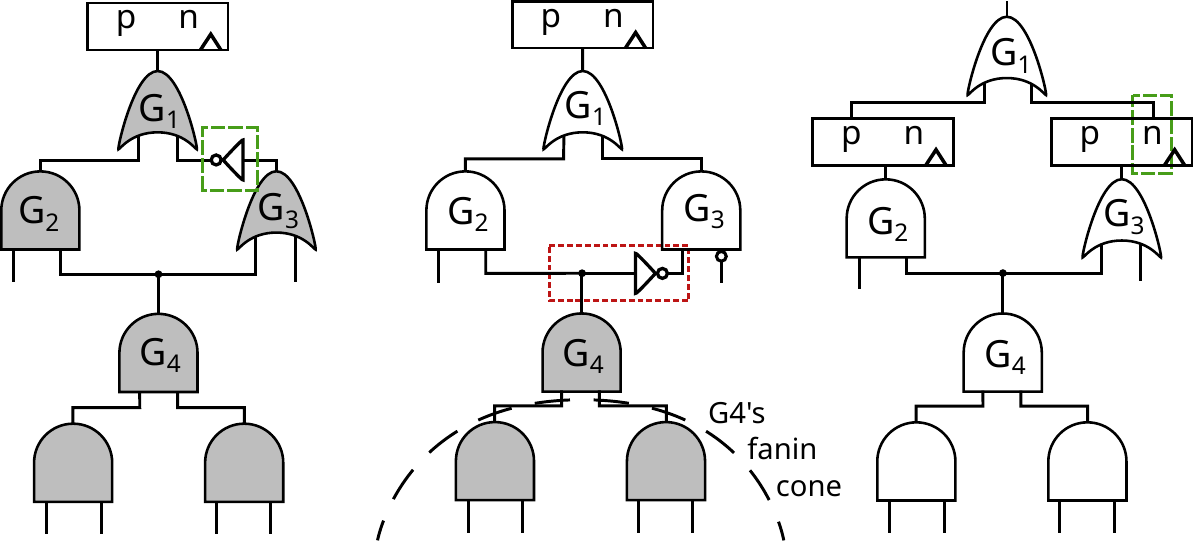}
    \caption{Example circuit before dual-rail translation. Gates that need duplication are shaded. Left---A naive dual-rail translation duplicates all gates. Middle---Applying De Morgan's rule to gate $G_3$ removes the need to duplicate it ($G_1$, $G_2$, and $G_3$ now appear unshaded). De-duplication stops at $G_4$ because both output polarities are required. Right---Repositioning DROC cells with a duplication-aware retiming strategy further de-duplicates $G_4$ and its fanin cone.}
    \label{fig:reconvergence}
\end{figure}

Figure~\ref{fig:reconvergence} shows an example gate-level circuit before dual-rail translation, with gates to be mapped to FA–LA pairs (i.e., duplicated) shaded. The left panel illustrates a naive approach in which every gate is duplicated. The middle panel shows the approach used in prior xSFQ synthesis work~\cite{xsfq-dac}: the netlist is traversed backward from the registers, and De Morgan's rule is applied to $G_3$, removing three previously duplicated gates (now $G_1$, $G_2$, and $G_3$ are unshaded). Duplication remains necessary at $G_4$ because both polarities are required, which in turn forces all gates in $G_4$'s fanin cone to be duplicated. The right panel demonstrates an alternative that eliminates duplication entirely, motivating the proposed minimum-area retiming strategy. 
By moving the DROC cell backward, the original inverter can be absorbed, removing conflicting-polarity paths. The extent to which the DROC cell can be moved depends on timing constraints; here, it is assumed that registers are only permitted at the inputs or output of $G_1$.

\subsection{Proposed Method}
\label{sec:proposed-retiming}
We propose a minimum-area retiming strategy for dual-rail circuits that balances the trade-off between register count and gate duplication. Minimizing register count (i.e., number of DROC cells) does not always reduce total cell or JJ count. Likewise, maximizing cell de-duplication can increase the register count, with each register (i.e., DROC cell) costing more JJs than either a single FA or LA cell.

The challenge stems from the interdependence of register placement and polarity assignment: repositioning the output registers of a pipeline stage can change optimal polarities, which in turn affects the placement of its input registers. Register repositioning may also require downstream adjustments in the circuit to meet timing constraints, further affecting placement and duplication costs.

We address this by formulating the problem as an integer linear program (ILP), outlined in Section~\ref{sec:ilp_formulation}. The ILP encodes the effects of register placement and polarity assignment on gate duplication through constraint-based rules, while the objective---minimizing total JJ count as the sum of register and duplication costs---maps directly to a linear cost function over the relevant variables. To the best of our knowledge, this is the first retiming solution for monotonic circuits that jointly optimizes sequential (register placement) and combinational (polarity assignment) logic.

Prior ILP-based methods~\cite{retiming,minaret,duplication-ilp} addressed register count or gate duplication separately but did not optimize their combined cost, which is critical here. Related work on two-phase domino logic used heuristics to strategically insert registers and break duplication cascades within a combinational block~\cite{minesota}. However, this approach does not generalize: (a) it only handles two-stage splits and assumes fixed fanin cones with static node weights, so it cannot support general retiming tasks; (b) it assumes that the combinational stage feeding the inserted registers becomes fully duplication-free, ignoring polarity conflicts deeper in the stage. For instance, in a circuit like Figure~\ref{fig:reconvergence}, it would keep a register at the output of $G_1$ because it cannot detect the polarity conflict at the preceding stage.

\subsection{ILP Formulation}\label{sec:ilp_formulation}
The ILP solver receives the original circuit, represented as an And-Inverter-Graph (AIG)~\cite{aig}, and outputs register placements and polarity assignments that minimize the combined cost of registers and gate duplication. The AIG consists of a set of nodes ($V$) and edges ($E$), where each node represents an AND gate. Each edge $(i,j)$ is assigned a binary value: $O_{i,j}=1$ indicates the presence of an inverter in the original graph, and $O_{i,j}=0$ indicates its absence. Additionally, a constant ($reg_{i,j} \in \mathbb{N}$) denotes the number of registers on that edge. Pre-retiming, all sets of registers---equal to the number of pipeline stages ($P$)---are placed at the primary inputs, and the $reg$ values on their outgoing edges are set to $P$.


\subsubsection{Variables}
\begin{itemize}[leftmargin=*]
  \item \textit{Retiming label:} For each node $v \in V$, the integer variable $r_v \in \mathbb{Z}$ indicates the number of registers to move across the node. Positive values indicate backward repositioning, while negative values indicate forward repositioning. 
  \item \textit{Polarity label:} For each node $v \in V$, the variable $p_v \in \{0, 1\}$ indicates whether De Morgan's law is applied at the node. If $p_v = 1$, the logical polarity inverts; if $p_v = 0$, it remains unchanged.
  \item \textit{Inversion label:} For each edge $(i,j) \in E$, the variable $x_{i,j} \in \{0,1\}$ indicates whether an inverter is present on the $(i,j)$ edge post-retiming. Its value, 1 for inverter and 0 otherwise, depends on the polarity labels of the adjacent gates.
  \item \textit{Duplication label:} For each node $v \in V$, the variable $d_v \in \{0, 1\}$ indicates whether the node requires duplication. Its value, 1 for duplication and 0 otherwise, depends on the retiming and polarity labels as determined by the constraint system.  
\end{itemize}

\subsubsection{Constraints}
\begin{itemize}[leftmargin=*]
    \item \textit{Non-negative register counts:} Post-retiming register counts must remain non-negative:\[\text{regs}'_{i,j} \geq 0 \quad \forall (i,j) \in E\] 
    \[ \text{regs}'_{i,j} = \text{regs}_{i,j} + r_j - r_i \quad \forall (i,j) \in E\] Here, $\text{regs}_{i,j}$ is the initial register count on the given edge, $r_j$ represents the registers added from the destination node $j$, and $r_i$ represents the registers removed from the source node $i$. 
  \item \textit{Clock period satisfaction:} Any path with delay exceeding the clock period $T$ must include at least one register:\[\text{regs}'_{\text{path}(i,j)} \geq 1 \quad \forall \text{ paths with } D_{\text{path}(i,j)} > T\] \[\text{regs}'_{\text{path}(i,j)} = \text{regs}_{\text{path}(i,j)} + r_j - r_i\] Here, $D_{\text{path}(i,j)}$ denotes the longest path delay, including both gate and fanout delays, and $\text{regs}_{\text{path}(i,j)}$ is a constant representing the original register count along any path between nodes $i$ and $j$. 

  
  \item \textit{Input/output preservation:} No registers may be moved across primary inputs (PI) or outputs (PO):\[r_v = 0 \quad \forall v \in \text{PI} \cup \text{PO}\]
  \item \textit{De Morgan's rule:} Applying De Morgan's rule at either endpoint of an edge toggles the presence of an inverter. If the initial inverter state is ($O_{i,j}$), the final inverter state is \[x_{i,j} = O_{i,j} \oplus p_i \oplus p_j\] To linearize this three-input XOR for ILP, we introduce an auxiliary variable $y_{i,j} \in \{0,1\}$ and rewrite the constraint as \[2y_{i,j} = x_{i,j} + O_{i,j} + p_i - p_j\]
  \item \textit{Duplication requirement:} Node $i$ requires duplication when any successor node $j \in \text{succ}(i)$ is reachable  without crossing a register ($\text{regs}'_{i,j}=0$) and either the edge carries an inverter or $j$ is itself duplicated. Formally, \[d_i = \exists j \in \text{succ}(i): \big[( \text{regs}'_{i,j} = 0) \land (x_{i,j} \lor d_j)\big]\] To linearize this condition, we introduce auxiliary variables $s_{i,j} \in \{0,1\}$ and $m_{i,j} \in \{0,1,2\}$, yielding: \[s_{i,j} M \ge \text{regs}'_{i,j}, \qquad s_{i,j} \le \text{regs}'_{i,j}, \qquad m_{i,j} \le 2 s_{i,j}, \]\[ m_{i,j} \le d_j + x_{i,j}, \qquad m_{i,j} \ge (d_j + x_{i,j}) - 2(1 - s_{i,j}),\] \[d_i M \ge \sum_{j \in \text{succ}(i)} \big(d_j + x_{i,j} - m_{i,j}\big)\] where $M$ is a sufficiently large constant.
\end{itemize}

\subsubsection{Objective Function}\label{sec:obj_function}

The ILP solver minimizes total area by balancing register count and gate duplication:
\[
\min \quad 
\alpha \sum_{v \in V} \Delta_{\text{reg}}(v)\, r_v 
\;+\;
\beta \sum_{v \in V \setminus (\text{PI} \cup \text{PO})} d_v,
\]
where $\alpha$ denotes the area cost per register,  $\beta$ is the area cost per duplicated gate (including interface overhead), and  $\Delta_{\text{reg}}(v)$ is a constant capturing the net register change induced when  retiming across node $v$, accounting for register-sharing opportunities~\cite{timing-book}.

\section{Cell Interface Redesign} 
\label{sec:bisquid}

\subsection{Existing Interfaces}
\subsubsection{Gate interconnection} In SCE circuits, gates are either abutted or connected via passive transmission lines (PTLs), with the latter being the standard in all place-and-route tools~\cite{coldflux, Nath2020, 10087349}. As with conventional transmission lines, maximizing signal integrity with PTLs requires optimized hardware interfaces. For this reason, dedicated PTL receivers and drivers are placed at every gate input and output, each costing 3 and 2~JJs, respectively~\cite{schindlerptl}. This results in an 8~JJ overhead for an xSFQ gate with two inputs and one output. 

\subsubsection{Fanout} 
The limited energy of an SFQ pulse introduces another interface overhead, that for fanout greater than one. Copying an SFQ by dividing it in two, without adding more energy, degrades circuit operating margins and can prevent loading cells from being driven. This is typically addressed with one or more splitter cells, which act like amplifiers. Each fanout-2 (FO2) splitter costs 3~JJs, excluding the PTL drivers and receivers at its I/O ports.


\subsubsection{Cost summary} For a cell library of 2-input, 1-output gates, one FO2 splitter is needed at the output of every cell~\cite{xsfq-dac}. This raises the average xSFQ cell interface cost to 13~JJs for FO2 with abutment between the xSFQ gate and splitter ($2\times$ PTL receivers at the gate inputs, $1\times$ splitter, and $2\times$ PTL drivers at the splitter outputs) or 18~JJs without abutment ($3\times$ PTL receivers at gate and splitter inputs, $1\times$ splitter, and $3\times$ PTL drivers at outputs). For reference, the logic of an xSFQ gate requires only 4~JJs. In other words, the cost of interfacing exceeds that of the logic by more than $3\times$. Note that this overhead is even higher for RSFQ designs due to the extra clock port: the interface cost per gate ranges from 21~JJs (with maximum abutment) to 31~JJs (without any abutment).


\begin{figure} [t]
    \centering
    \includegraphics[width=0.5\linewidth]{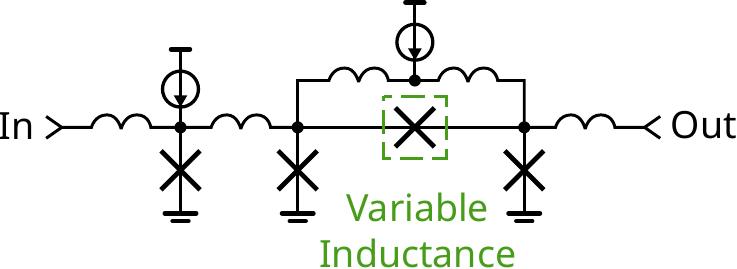}
    \vspace{-2mm}
    \caption{Proposed amplifier schematic, featuring a variable-inductance path for higher fanout and PTL driving capability than conventional splitters and PTL interfaces.}
    \label{fig:bisquid-schem}
\end{figure}

\subsection{Proposed Interface}
\label{sec:interface}
We introduce a new amplifier cell to minimize the dominant interface costs. Our design is inspired by the BiSQUID, a device that has a linear magnetic flux-to-voltage response and is typically used for magnetic field detection~\cite{Kornev_2009}. However, there is no prior record of use for amplification in digital SFQ circuits.

\begin{table}[b]
\vspace{-2mm}
\centering
\caption{JJ count and delay metrics for proposed LA, FA, DROC, and SPLIT cells versus xSFQ-DAC61~\cite{xsfq-dac}; for DROC, left/right of the slash indicate clock-to-Q$_p$/Q$_n$ delays.}
\label{tab:cells}
\resizebox{0.95\columnwidth}{!}{
\begin{tabular}{|l|l|r | rrrr|}
\hline
&&&\multicolumn{4}{c|}{\textbf{Delay (ps)}}\\\cline{4-7}
\textbf{Cell} & \textbf{Library} & \textbf{\#JJs} & \textbf{FO-1} & \textbf{FO-2 } & \textbf{FO-3 } & \textbf{FO-4} \\
\hline

\multirow{2}{*}{LA} &  xSFQ-DAC61 & 12  & 19.9 &  --- &  --- &  --- \\ \cline{2-7}
                     & Proposed &  7  &  7.9 &  8.7 &  9.5 & 10.4 \\
\hline 

\multirow{2}{*}{FA} & xSFQ-DAC61 & 12 & 24.7 &  --- &  --- &  --- \\ \cline{2-7}
                    & Proposed &  7 &  7.5 &  8.3 &  9.2 & 10.2 \\
\hline

 \multirow{1}{0.5cm}{DROC} & xSFQ-DAC61 & 23 & 18.0/21.5 &  --- &  --- &  --- \\ \cline{2-7}
                         & Proposed & 19 &  9.0/13.3 &  9.9/14.0 & 10.7/14.8 & 11.4/15.5 \\
\hline


\multirow{2}{*}{SPLIT} & xSFQ-DAC61 & 10 &  --- & 19.7 &  --- &  --- \\ \cline{2-7}
                       & Proposed &  4 &  3.5 &  4.4 &  5.4 &  6.4 \\
\hline
\end{tabular}
}
\end{table}

\begin{table*}[t]
\centering
\caption{JJ count comparison between the baseline (Section~\ref{baseline}) and the proposed duplication-aware retiming. Benchmarks are ISCAS85 circuits, standard in SCE logic synthesis. All experiments assume the xSFQ-DAC61 cell library to isolate retiming benefits from those of the integrated xSFQ-amplifier cells. Strict and relaxed timing constraints are in Section~\ref{retiming-results}.}
\label{tab:pipelined_default}
\resizebox{\textwidth}{!}{
\Huge
\begin{tabular}{|l|r|r|R|r|r|R|r|r|R|r|r|R|r|r|R|r|r|R|}
\hline
& \multicolumn{6}{c|}{\textbf{2 Pipeline Stages}} & \multicolumn{6}{c|}{\textbf{4 Pipeline Stages}} & \multicolumn{6}{c|}{\textbf{6 Pipeline Stages}} \\
\cline{2-19}
\multicolumn{1}{|c|}{\textbf{Circuit}}& \multicolumn{3}{c|}{\textbf{Strict Timing}} & \multicolumn{3}{c|}{\textbf{Relaxed Timing}} & \multicolumn{3}{c|}{\textbf{Strict Timing}} & \multicolumn{3}{c|}{\textbf{Relaxed Timing}} & \multicolumn{3}{c|}{\textbf{Strict Timing}} & \multicolumn{3}{c|}{\textbf{Relaxed Timing}} \\
\cline{2-19}
\textbf{} & \textbf{Baseline} & \textbf{Proposed} & \textbf{Red.} & \textbf{Baseline} & \textbf{Proposed} & \textbf{Red.} & \textbf{Baseline} & \textbf{Proposed} & \textbf{Red.} & \textbf{Baseline} & \textbf{Proposed} & \textbf{Red.} & \textbf{Baseline} & \textbf{Proposed} & \textbf{Red.} & \textbf{Baseline} & \textbf{Proposed} & \textbf{Red.} \\
\hline
c1355 & 14,802 & 14,802 & 0\% & 14,568 & 12,728 & 13\% & 15,152 & 14,936 & 1\% & 16,351 & 13,522 & 17\% & 17,663 & 15,154 & 14\% & 17,424 & 14,800 & 15\% \\
\hline
c1908 & 10,461 & 10,217 & 2\% & 11,248 & 9,877 & 12\% & 11,390 & 10,544 & 7\% & 11,326 & 10,217 & 10\% & 12,558 & 11,811 & 6\% & 12,606 & 11,493 & 9\% \\
\hline
c2670 & 17,967 & 14,786 & 18\% & 18,430 & 14,454 & 22\% & 19,504 & 17,983 & 8\% & 18,762 & 16,799 & 10\% & 23,285 & 22,178 & 5\% & 23,113 & 21,719 & 6\% \\
\hline
c3540 & 27,930 & 27,325 & 2\% & 27,737 & 24,580 & 11\% & 26,063 & 24,894 & 4\% & 26,148 & 23,041 & 12\% & 27,106 & 25,437 & 6\% & 26,864 & 25,094 & 7\% \\
\hline
c432 & 3,429 & 3,429 & 0\% & 4,384 & 3,455 & 21\% & 3,789 & 3,789 & 0\% & 3,486 & 3,486 & 0\% & 5,743 & 5,763 & 0\% & 5,339 & 5,339 & 0\% \\
\hline
c499 & 14,782 & 14,782 & 0\% & 15,058 & 12,728 & 15\% & 15,626 & 15,078 & 4\% & 15,421 & 13,596 & 12\% & 17,565 & 15,077 & 14\% & 17,315 & 14,804 & 15\% \\
\hline
c5315 & 43,148 & 38,627 & 10\% & 43,391 & 37,906 & 13\% & 48,512 & 43,290 & 11\% & 49,321 & 42,272 & 14\% & 54,092 & 49,995 & 8\% & 54,329 & 48,083 & 11\% \\
\hline
c6288 & 80,509 & 79,274 & 2\% & 81,335 & 78,567 & 3\% & 79,450 & 77,957 & 2\% & 80,255 & 76,834 & 4\% & 79,148 & 76,465 & 3\% & 80,473 & 75,708 & 6\% \\
\hline
c7552 & 54,810 & 47,327 & 14\% & 55,322 & 46,547 & 16\% & 58,050 & 47,015 & 19\% & 58,187 & 46,396 & 20\% & 59,670 & 50,688 & 15\% & 60,995 & 50,191 & 18\% \\
\hline
c880 & 9,634 & 8,128 & 16\% & 10,038 & 8,033 & 20\% & 10,925 & 9,961 & 9\% & 10,447 & 9,472 & 9\% & 12,637 & 11,923 & 6\% & 12,645 & 11,841 & 6\% \\
\hline
\textbf{Geomean} & --- & --- & 7\% & --- & --- & 15\% & --- & --- & 7\% & --- & --- & 11\% & --- & --- & 8\% & --- & --- & 9\% \\
\hline
\end{tabular}
}
\end{table*}
\subsubsection{Key characteristics \& functionality}
The proposed amplifier differs from the PTL receivers, PTL drivers, and splitters discussed above. Instead of passing an SFQ from input to output along a fixed-inductance path, it provides a path between two JJs whose inductance changes nonlinearly as the SFQ propagates (Figure~\ref{fig:bisquid-schem}).

At the start of the SFQ pulse, the cell behaves like a much larger JJ ($\sim2.5\times$ the size of other JJs in the circuit) due to the low inductance between the two JJs. Normally, a single SFQ pulse does not have enough energy to switch such a large JJ without preceding amplification stages, causing the SFQ to bleed to ground. In this cell, however, the inductance increases at the peak of the SFQ pulse, effectively turning the cell into two separate, smaller JJs. Combined with the subsequent redirection of bias and SFQ current into the output, this allows the SFQ pulse to propagate through the cell and drive subsequent fanout and PTLs---at a cost of just 4 JJs. 
 
The same varying-inductance path also improves circuit reliability compared to prior amplifiers. In near-overbias conditions, it redirects current away from the JJ closest to switching. In near-underbias conditions, it slows the bias decrease for the most sensitive JJ. This enables operation at lower-than-typical bias points.


\subsubsection{Integration with xSFQ}
\label{sec:integration}
Next, we integrate the proposed amplifier at the outputs of the three xSFQ cells: FA, LA, and DROC. This removes the need for conventional PTL receivers at the inputs, PTL drivers at the outputs, and splitters for FO2–FO4 fanouts. With JJ-sharing optimizations~\cite{icranking}, the cells' output JJs are merged with the amplifier's input JJ, saving one junction. Consequently, the cumulative cost of the new xSFQ FA and LA cells is reduced to 7 JJs (4 for logic + 3 for the integrated amplifier), compared to 17/22 JJs (with/without abutment) for FO2 and 27/42 JJs for FO4.

The DROC cell benefits similarly, achieving FO1–FO4 while replacing PTL receivers, drivers, and splitters with the amplifier, at the cost of 3 JJs per output. Fanouts beyond FO4 can be realized by chaining the amplifier as a standalone element (SPLIT), adding one FO2, FO3, or FO4 at a time.


\subsection{Cell Library}
Table~\ref{tab:cells} summarizes JJ count and propagation delays for the integrated xSFQ-amplifier cells. For reference, it also includes JJ counts and timing data from the most recent xSFQ cell library~\cite{xsfq-dac}, which we refer to as xSFQ-DAC61 for brevity. In that library, the SPLIT is a conventional FO2 splitter and the FA, LA, and DROC cells only support FO1; hence, the dashes in their higher-fanout entries. By contrast, every proposed cell natively supports up to FO4, with the new SPLIT cell also providing additional FO4 fanout. In the case of DROC, the cost of splitting in the clock network is excluded.

The proposed cells reduce JJ count by 
1.7–3.1$\times$ for FO2 and 2.8–6.0$\times$ for FO4 
(detailed breakdown in Section~\ref{sec:interface})  and cut cell delay by 2–4.5$\times$. The speedup is due to fewer JJs on the critical paths, even before accounting for the fanout overhead in xSFQ-DAC61.

\section{Evaluation} 
\label{sec:eval}
\subsection{Duplication-Aware Retiming}
\label{sec:eval1}



\subsubsection{Baseline} 
\label{baseline}We evaluate duplication-aware retiming using JJ count as the main metric. While xSFQ-DAC61 could serve as a baseline, it supports only ABC's default retiming. To provide a stronger comparison point, we (a) add minimum-register retiming~\cite{minaret} and (b) replace its heuristic polarity assignment~\cite{domino-graph} with an ILP-based method~\cite{duplication-ilp}. These enhancements better reflect the state of the art for monotonic circuits (Section~\ref{sec:proposed-retiming}). In the remainder of the paper, ``baseline'' refers to xSFQ-DAC61 with these enhancements.

\subsubsection{Methodology} Each benchmark circuit is first synthesized in ABC~\cite{abc} for AIG optimization. The resulting AIG~\cite{aig} is then parsed to compute fanout-dependent delays arising from splitter insertions, providing the node delays used in the retiming graph. We then construct the ILP variables and constraints described in Section~\ref{sec:3} and solve the ILP using Gurobi~12.0.3 with 16 threads and a one-hour timeout. All ISCAS85~\cite{iscas85} circuits, the standard benchmark for SCE logic synthesis, complete well within this time limit.

Using the ILP solution, we generate Verilog netlists, where each AIG node is mapped to an FA cell, LA cell, or FA–LA pair, registers are mapped to DROC cells, and SPLIT cells are inserted as needed. ABC is then used to verify equivalence against the original benchmark and extract both critical-path delay and total JJ count.  

\begin{figure}[bht]
    \centering
    \includegraphics[width=0.76\linewidth]{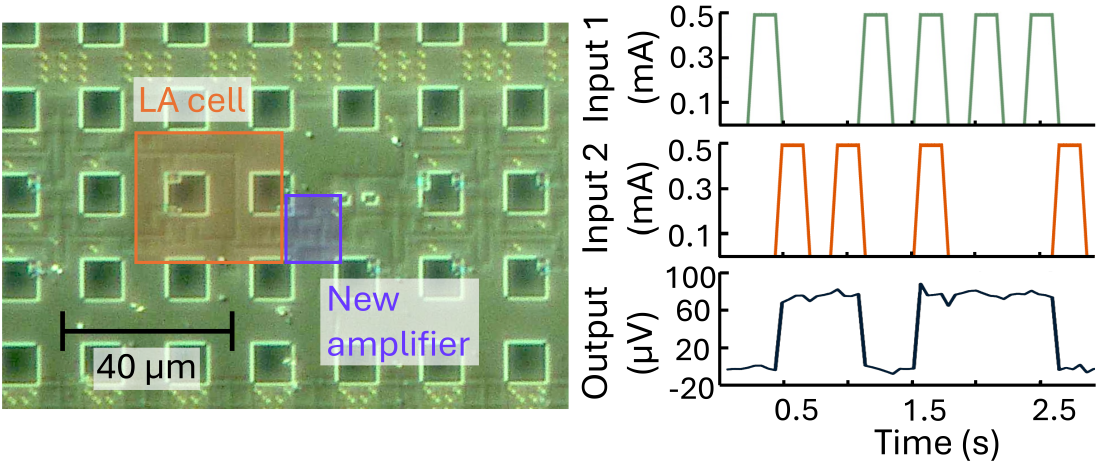}
    \caption{Left---Photo of the fabricated circuit with LA cell and integrated amplifier driving FO2. Right---Measured output voltage transitions confirm correct operation.}
    
    \label{fig:protophoto}
\end{figure}

\subsubsection{Results} 
\label{retiming-results}

Table~\ref{tab:pipelined_default} reports the post-retiming JJ counts. All experiments use the xSFQ-DAC61 cell library (including PTL interface costs as reported in the original paper~\cite{xsfq-dac}) to isolate the benefits of duplication-aware retiming from those of the new integrated xSFQ-amplifier cells. We sweep the pipeline depth $P$ from two to six stages (in steps of two) and evaluate two target periods for each $P$: (a) a \textit{strict} period, defined as the minimum delay achievable without increasing $P$ to $P{+}1$, and (b) a \textit{relaxed} period, defined as the maximum delay that is not supported by $P{-}1$. These two settings bound the practical clock period range for each pipeline depth.


In summary, JJ count reduction across ISCAS85 circuits averages from 7–15\% and reaches a maximum of 22\%. Gains are consistently higher under the relaxed timing target, where the ILP has more freedom for register repositioning. In several cases, improvements occur in both throughput and JJ count. For example, \textit{c3540} requires 35,146 JJs as a single-cycle design ($P=1$); after 4-stage pipelining ($P=4$) and retiming, it uses only 23,041 JJs, representing a 34\% reduction and a clear Pareto improvement. More data for $P=1$ are in Table~\ref{tab:combinational_baseline} under the xSFQ-DAC61 column.

\begin{table}[t]
\centering
\small
\setlength{\tabcolsep}{3pt}
\caption{JJ count comparison between xSFQ-DAC61 and the proposed library across single-cycle ISCAS85/EPFL circuits.}
\label{tab:combinational_baseline}
\resizebox{0.44\textwidth}{!}{
\begin{tabular}{|l|r|r|r|r|r|r|R|} 
\hline
\textbf{Circuit} & \multicolumn{3}{c|}{\textbf{xSFQ-DAC61~\cite{xsfq-dac}}} & \multicolumn{3}{c|}{\textbf{Proposed Library}} & \\
\cline{2-4} \cline{5-7} 
& \textbf{Gates} & \textbf{Split.} & \textbf{\#JJ} & \textbf{Gates} & \textbf{Split.} & \textbf{\#JJ} & \textbf{Improv.} \\
\hline
c880 & 460 & 386 & 9,380 & 460 & 28 & 3,332 & 2.82$\times$ \\ 
\hline
c1908 & 498 & 457 & 10,546 & 498 & 54 & 3,702 & 2.85$\times$ \\ 
\hline
c499 & 678 & 628 & 14,416 & 678 & 84 & 5,082 & 2.84$\times$ \\ 
\hline
c3540 & 1633 & 1555 & 35,146 & 1633 & 256 & 12,455 & 2.82$\times$ \\ 
\hline
c5315 & 1911 & 1771 & 40,642 & 1911 & 290 & 14,537 & 2.80$\times$ \\ 
\hline
c7552 & 2560 & 2278 & 53,500 & 2560 & 197 & 18,708 & 2.86$\times$ \\ 
\hline
int2float & 229 & 214 & 4,888 & 229 & 40 & 1,763 & 2.77$\times$ \\ 
\hline
dec & 304 & 544 & 9,088 & 304 & 128 & 2,640 & 3.44$\times$ \\ 
\hline
priority & 847 & 601 & 16,174 & 847 & 0 & 5,929 & 2.73$\times$ \\ 
\hline
sin & 9913 & 9890 & 217,856 & 9913 & 1543 & 75,563 & 2.88$\times$ \\ 
\hline
cavlc & 712 & 703 & 15,574 & 712 & 151 & 5,588 & 2.79$\times$ \\ 
\hline
\end{tabular}
}
\end{table}




\subsection{Cell Interface Redesign}\label{sec:eval_interface}

\subsubsection{Functional validation through simulation}
All integrated xSFQ-amplifier cells were designed and simulated in Cadence Virtuoso using MITLL's SFQ5ee 10 kA/cm\textsuperscript{2} device models and verified to have global bias margins of at least $\pm15\%$. To assess global margins, each cell was simulated while both driving and being loaded by all valid combinations of fanouts (from 1 to 4) and integrated xSFQ-amplifier cells.

PTLs were modeled as Nb striplines with 4.5~$\mu$m widths and 50~$\mu$m lengths, representative of typical interconnect distances between cells~\cite{interconnect}. Propagation delays (FA, LA, SPLIT) and clock-to-Q delays (DROC) were extracted from the timing between input-flux and output-flux transitions of the cell under test, averaged over all driving/loading combinations. Results are summarized in Table~\ref{tab:cells}.




\subsubsection{Functional validation through fabrication}
To experimentally validate the integrated xSFQ-amplifier design, we fabricated an LA cell with FO2 in MITLL's SFQ5ee process and measured it at $4.2 \text{K}$. Figure~\ref{fig:protophoto} shows the physical prototype (left) and related measurements (right). Input current and output voltage transitions indicate logic 1; all transitions demonstrate correct operation.

\subsubsection{Results}
Table~\ref{tab:combinational_baseline} reports the impact of the new cell library on JJ counts for the same ISCAS85~\cite{iscas85} and EPFL~\cite{epfl} combinational circuits as in xSFQ-DAC61. To isolate the benefits of the redesigned cell interfaces from retiming, we set $P=1$. For both xSFQ-DAC61 and the proposed library, PTL interface costs are included in the analysis (the original xSFQ-DAC61 excluded these overheads).

The results show that the new library reduces JJ count by 2.7–3.4$\times$ compared to xSFQ-DAC61, with splitter counts dropping sharply due to the cells' integrated fanout capability; reaching zero for the \textit{priority} benchmark. For reference, xSFQ-DAC61 achieved an average 4.5$\times$ JJ reduction over prior RSFQ synthesis results~\cite{pbmap} before accounting for PTL and clock network overheads.

\subsection{Combined Improvements}\label{sec:alltogether}
\begin{figure}[h]
    \centering
    \includegraphics[width=1\linewidth]{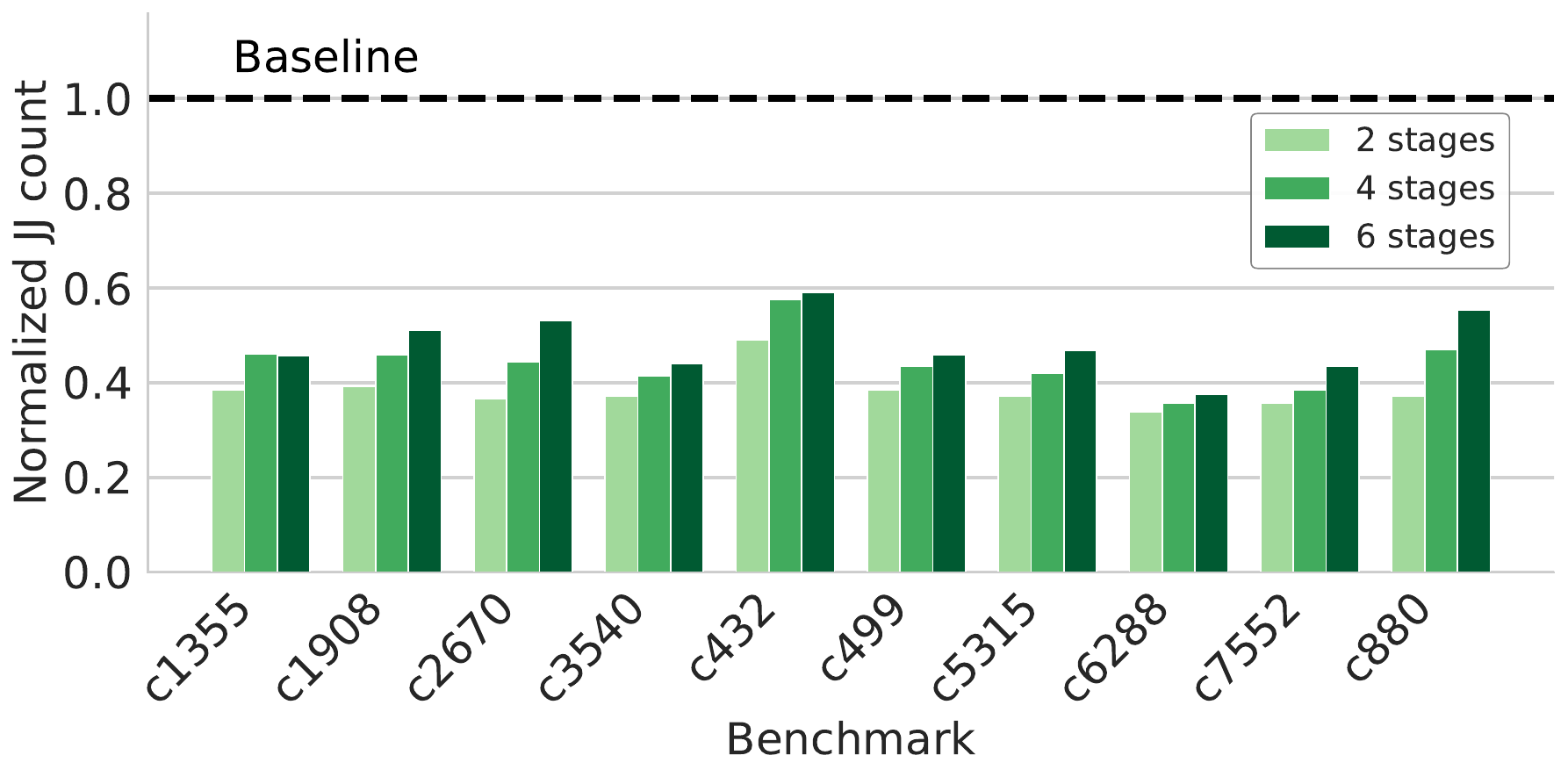}
    \caption{Results for the proposed duplication-aware retiming algorithm and cell library on ISCAS85 circuits across 2, 4, and 6 pipeline stages. JJ counts are normalized to the respective baseline (enhanced xSFQ-DAC61, Section~\ref{baseline}), showing reductions of 41–66\%.}
    \label{fig:area_comparison}
\end{figure}

The final step in our evaluation quantifies the combined impact of duplication-aware retiming and the redesigned cell interfaces. Figure~\ref{fig:area_comparison} reports the JJ count reduction for pipelined circuits ($P$ from 2 to 6 in increments of 2), normalized to the counts under the xSFQ-DAC61 cell library and the enhanced baseline flow described in Section~\ref{sec:eval1}. The proposed approach reduces JJ count by 41–66\%.

Differences relative to the gains in Table~\ref{tab:combinational_baseline} stem from pipelining: introducing DROC cells, the base designs of which use roughly 3$\times$ more JJs than FA/LA cells, may limit the percentage savings. Still these designs establish a new Pareto frontier, achieving both shorter critical paths and lower JJ count than the best-to-date single-cycle implementations~\cite{xsfq-dac}.



\section{Conclusion} 
\label{sec:concusion}

We present a duplication-aware retiming method combined with a cell interface redesign for SFQ circuits. To our knowledge, this is the first retiming solution for monotonic circuits that jointly optimizes sequential and combinational logic, as well as the most compact SFQ cell library to date. The approach achieves 1.7–3.4$\times$ JJ count reduction over the prior state of the art for dual-rail SFQ, depending on the degree of pipelining, translating to significant improvements over traditional SFQ designs.


\newpage

\bibliographystyle{ACM-Reference-Format}
\bibliography{refs}

\end{document}